\documentclass[preprint,aps,amssymb,showpacs,superscriptaddress,nofootinbib]{revtex4-2}

\usepackage{graphicx}
\usepackage{dcolumn}   
\usepackage{bm}   
\usepackage{amssymb}
\usepackage{color}
\usepackage{amsmath}
\usepackage{mathrsfs}
\usepackage{latexsym}
\usepackage{natbib}    
\usepackage[toc,page]{appendix}
\usepackage{braket}
\newcommand{\del}{\partial}

\begin{document}

\title{Galactic Spacetime Solutions with a Varying Newton's Coupling}

\author{Rounak Chakraborty}
\email{rounakpapai@kgpian.iitkgp.ac.in}
\affiliation{Department of Physics, Indian Institute of Technology Kharagpur, Kharagpur-721302, INDIA}

\author{Sandipan Sengupta}
\email{sandipan@phy.iitkgp.ac.in}
\affiliation{Department of Physics, Indian Institute of Technology Kharagpur, Kharagpur-721302, INDIA}

\begin{abstract}

We find a new family of galactic metrics corresponding to flat rotation curves at the outer radii. 
These are vacuum solutions to a gravity theory where the Newton's coupling varies mildly in space. 
The effective `mass', whose origin is purely geometric, receives a negative non-baryonic contribution. 
The angle of deflection of a light ray propagating in this geometry is found to be diminished rather than enhanced compared to the Einsteinian bending, the effect being highly suppressed though. Hence, these spacetimes are observationally dintinguishable from other geometric or `dark matter' based alternatives invoked to explain mass discrepancies in galaxies.

\end{abstract}


\maketitle
\newpage


\section{Introduction}

The observations regarding the virial mass discrepancy for galaxy clusters \cite{zwicky} along with the large distance profile of the circular velocity of test particles in spiral galaxies \cite{rubin,*rubin1,roberts} bring in intriguing implications. These suggest that Einstein gravity as it is with standard matter coupling might not be completely satisfactory as the underlying theory at galactic scales or beyond. The essence of the problem may be summarized in a no-go result, that is, pure general relativity cannot provide a dynamical explanation to the observed flatness of rotation curves far away from the centre of spiral galaxies either in an exact or approximate form. Since the earliest observations in this context, there have been various attempts to construct dynamical models to this end within formulations beyond general relativity. These are motivated from both theoretical and phenomenological perspectives
 \cite{milgrom,*milgrom1,bek,mannheim,*mannheim1,sanders,nuc,bharadwaj,
 harko,sobouti,moffat,vagnozzi,seng}.

Here, we shall explore the question as to whether a small (practically almost insignificant) spatial variation of the gravitational coupling could dynamically generate spacetime geometries with asymptotically flat galactic rotation curves. The idea of variable couplings in a gravity Lagrangian has a long history, early formulations of which was mainly due to the works of Jordan and Brans-Dicke \cite{jordan,brans}. Our investigation here is based on a four-dimensional gravity action that is first order in form and linear in curvature, explored in refs.\cite{seng,sengcosm}. Here we study the torsional phase of the vacuum equations of motion, not analyzed earlier, in the context of static spherically symmetric spacetimes. We find a new set of exact solutions for galactic spacetimes. These are shown to correspond to constant rotational velocities of test particles at a large radii, far away from  luminous matter. 

To emphasize, our approach here is purely geometric. Matter-coupling, not relevant for the problem addressed in this article, could nevertheless be introduced exactly as in standard gravity theory.

The analysis of the deflection of a light ray in this geometry reveals a new effect. The angle of gravitational bending is found to be diminished compared to the Einsteinian angle, an effect that has no analogue in the standard modified gravity or `cold dark matter' (CDM) models. This modification, however, is practically very small.

We begin our analysis in Sec-II with a brief review of the gravity Lagrangian and the associated equations of motion. The static, spherically symmetric case and the Newtonian solutions are discussed next. In Sec-III, we find the galactic spacetime solutions, elucidating in detail on their critical features relevant to the context here. Subsequently, we study the deflection of light in these spacetimes, with possible observational imports. The final section contains a few relevant remarks. Additional details regarding the bending angle in a finite halo geometry joined to a Schwarzschild spacetime is provided at the Appendix.

\section{The Lagrangian formulation}

Our starting point is the following action principle for gravity in four dimensions \cite{seng,sengcosm}:
\begin{eqnarray}\label{S}
S[\xi,e,\omega,\psi]~=~\int d^4 x ~ e \big[\xi e^{\mu}_{I} e^\nu_J {R}_{\mu\nu}^{~~IJ}(\omega) ~+~{\cal L}_m (\psi)\big]
\end{eqnarray}
where $\xi$ is a scalar, $e_\mu^I$ are the tetrad, $\omega_\mu^{~IJ}$ are the spin-connection defining the field-strength as: $R_{\beta\rho}^{~~LM}(\omega)=\del_{[\beta} \omega_{\rho]}^{~LM}+\omega_{[\beta}^{~LK}\omega_{\rho]K}^{~~~M}$.  The last term at the RHS denotes the matter Lagrangian which depends on the tetrad and the matter fields 
$\psi$ only. For the context and our purpose here, we would ignore the cosmological term.

The following set of three equations of motion are obtained by varying (\ref{S}) with respect to the independent fields $\xi,e_\mu^I,\omega_\mu^{~IJ}$:
\begin{eqnarray}\label{eom1}
&& \delta\xi:~~ R(\omega)=0,\nonumber\\
&&\delta e:~~\xi\big[R_{\alpha\beta}^{~~IJ}(\omega)e^\beta_J-\frac{1}{2}e_{\alpha}^I R(\omega)\big]= 8\pi e^{\beta I}T_{\alpha\beta},\nonumber\\
&&\delta \omega:~~ D_\alpha(\omega)\big[\xi e e^{[\alpha}_I e^{\beta]}_J\big]=0,
\end{eqnarray}
where $R_{\alpha\beta}^{~~IJ} (\omega)$ is the field-strength with torsion in general, $8\pi T_{\alpha\beta}\equiv -\frac{\kappa}{e}e_{\beta I}\frac{\del {\cal L}_m}{\del e^\alpha_I}$ is the matter stress-energy tensor and $D_\mu(\omega)$ is the gauge-covariant derivative with respect to the connection. The solution to the connection equation above is given by: $\omega_\mu^{~IJ}=\bar{\omega}_\mu^{~IJ}(e)+K_\mu^{~IJ}$ , where $\bar{\omega}_\mu^{~IJ}(e)$ is the connection without torsion and the contortion reads: 
\begin{eqnarray}\label{K}
 K_\mu^{~IJ}=\frac{1}{2\xi}e^{\sigma [J}e_\mu^{I]}\del_\sigma \xi
 \end{eqnarray}

The static, spherically symmetric solutions for $\xi=0$, encoding a special non-Einsteinian vacuum phase of the above theory, has been studied in ref.\cite{seng} with vanishing torsion. The cosmological consequences of a generalized version of (\ref{S}), associated with time-varying Newton's and cosmological couplings with $\xi\neq 0$ have been elucidated in ref.\cite{sengcosm}. Here, we shall investigate the possible solutions with a spatially varying $\xi$ in the context of static, spherically symmetric spacetimes. The results along with  their intriguing implications are discussed next.

\section{Static, spherically symmetric torsional vacuum}

In vacuum, we have $T_{\alpha\beta}=0$. For 
$\xi\neq 0$ phase, the equations of motion then reduce to:
\begin{eqnarray}\label{Req}
R_{\alpha\beta}(\bar{\omega}+K)=0
\end{eqnarray}
where the left hand side above denotes the Ricci tensor with torsion defined through eq.(\ref{K}). Note that standard Einstein gravity with a vanishing torsion is recovered for $\xi=const$ ($K_\mu^{~IJ}=0$).

Spherically symmetric static solutions to these equations of motion correspond to a radially varying gravitational coupling $\xi=\xi(r)$ and a line element of the form ($ G=1=c$):
\begin{eqnarray*}
ds^2=-f(r)dt^2+g(r) dr^2+r^2(d\theta^2+\sin^2 \theta d\phi^2),
\end{eqnarray*}
Using the expression (\ref{K}) for contortion, the non-vanishing connection components with torsion are given by:
\begin{eqnarray*}
    \omega_t^{\hspace{1.5mm} 01} &=& \frac{1}{2}\Big(\frac{f}{g}\Big)^{\frac{1}{2}}\Big[\frac{f^\prime}{f} +  \frac{\xi^\prime}{\xi} \Big],~
    \omega_\theta^{\hspace{1.5mm} 12} = -g^{-\frac{1}{2}}\Big[1 + \frac{r}{2} \frac{\xi^\prime}{\xi}\Big],\\
    \omega_\phi^{\hspace{1.5mm} 31} &=& g^{-\frac{1}{2}}\Big[1 + \frac{r}{2} \frac{\xi^\prime}{\xi}\Big] \sin{\theta},~
    \omega_\phi^{\hspace{1.5mm} 23} = -\cos{\theta}.
\end{eqnarray*}
With these, the field equations eq.(\ref{Req}) reduce to a set of three independent equations owing to the nontrivial diagonal Ricci tensor components:
\begin{eqnarray}\label{tt-eqn}
    &&\Bigg[ \Bigg(\frac{f}{g}\Bigg)^{\frac{1}{2}} \Bigg( \frac{f^\prime}{f} + \frac{\xi^\prime}{\xi} \Bigg) \Bigg]^\prime  ~+~ \frac{2}{r}\Bigg(\frac{f}{g}\Bigg)^{\frac{1}{2}} \left( \frac{f^\prime}{f} + \frac{\xi^\prime}{\xi} \right) \left( 1+\frac{r}{2} \frac{\xi^\prime}{\xi} \right) ~=~ 0 , \\
    &&   \left[\Bigg(\frac{f}{g}\Bigg)^{\frac{1}{2}} \left( \frac{f^\prime}{f} + \frac{\xi^\prime}{\xi} \right) \right]^\prime ~+~ \frac{4f^{\frac{1}{2}}}{r} \Bigg[g^{-\frac{1}{2}} \left( 1+\frac{r}{2} \frac{\xi^\prime}{\xi}\right)\Bigg]' ~ =~ 0 ,\label{rr-eqn} \\
    && \left( 1+\frac{r}{2} \frac{\xi^\prime}{\xi}\right) \left[ \frac{f^\prime}{f} + \frac{\xi^\prime}{\xi} \right] ~+~2 g^{\frac{1}{2}} \Bigg[g^{-\frac{1}{2}}\left( 1+\frac{r}{2} \frac{\xi^\prime}{\xi}\right)\Bigg]^\prime~ -~ \frac{2}{r} \left[ g- \left( 1+\frac{r}{2} \frac{\xi^\prime}{\xi}\right)^2 \right]~ =~ 0~. \label{thetathetaeqn}
\end{eqnarray}
Here, `$\prime$' denotes the derivative with respect to $r$. Note that using equations (\ref{tt-eqn}) and (\ref{rr-eqn}), we obtain the following equation for $\xi$:
\begin{eqnarray}
\frac{2\left( 1+\frac{r}{2} \frac{\xi^\prime}{\xi}\right)'}{\left( 1+\frac{r}{2} \frac{\xi^\prime}{\xi}\right)}= \frac{f^\prime}{f} +\frac{g^\prime}{g}+ \frac{\xi^\prime}{\xi} 
\end{eqnarray}
This second order equation may be integrated, leading to the following result:
\begin{eqnarray}\label{fg}
\frac{1}{r\xi^{\frac{1}{2}}}=C+\bar{C}\int dr~\Bigg[\frac{(fg)^{\frac{1}{2}}}{r^2}\Bigg]
\end{eqnarray}

For the Newtonian class $(f(r)g(r)=1)$, we find the following general solution after redefining the integration constants:
\begin{eqnarray}
    \xi(r) = \frac{\xi_0}{(1-C r)^2}
\end{eqnarray}
Using this in \eqref{thetathetaeqn} we obtain:
\begin{eqnarray}
    f(r) = \frac{1}{g(r)} = (1- C r)^2 \left[ 1 + \frac{B(1-Cr)}{r}  \right] \label{schwtypesol}
\end{eqnarray}
This singular solution reduces to the Schwarzschild geometry when $C = 0$, i.e. $\xi = \xi_0$, corresponding to Einstein's gravity in vacuum. 

We do not explore the Newtonian sector any further here. The non-Newtonian solutions, which are the main subject of this article, are discussed next.

\section{Non-Newtonian solutions for galactic spacetimes}
Here, our goal is to find exact metric solutions to the equations of motion corresponding to galactic spacetimes.
It is well-known that such spacetimes are essentially non-Newtonian. To this end, let us look for solutions of the form: $f(r) g(r) = 1+ \delta(r)$ where $|\delta| << 1$.
To be specific, we may choose a convenient parametrization \cite{sobouti,seng} of the behaviour above as :
\begin{eqnarray}
    f(r) g(r) = \left( \frac{r}{R} \right)^{2\alpha} ,
\end{eqnarray}
where the length scale $R$ is required to preserve the dimensionless character of $f$ and $g$. It gets fixed in terms of the radial distance at which this metric could be matched with a Newtonian metric (e.g. Schwarzschild). The family of solutions is paramerized by $\alpha << 1$, whose interpretation would emerge from the subsequent analysis. 

The general solution for $\xi(r)$ in this case is obtained using eq.(\ref{fg}) as:
\begin{eqnarray}
    \xi(r) = \frac{1}{\left[C_1\left(\frac{r}{R}\right)^\alpha - C_2 \hspace{1mm} \left(\frac{r}{R}\right)\right]^2}, \label{xi-soln}
\end{eqnarray}
$C_1,~C_2$ being the redefined integration constants. For the above solution to be practically relevant, $\phi(r)$ must be a slowly varying function. This is possible provided: $C_2=0$.

Inserting the slowly varying solution for $\xi(r)$ above into eq.(\ref{thetathetaeqn}), we find the exact vacuum solution for the metric as:  
\begin{eqnarray}
    f(r) = \left( \frac{r}{R} \right)^{2\alpha}g^{-1}(r) =   \frac{\left( \frac{r}{R} \right)^{2\alpha}}{(1-\alpha)^2} - C \left( \frac{r}{R} \right)^{-1+3\alpha}  \label{g-metric}
\end{eqnarray}
Note that, for $\alpha = 0$, which corresponds to $\xi = const$, this reduces to the exterior Schwarzchild metric in vacuum, provided we identify the integration constant with the baryonic mass at the interior: $\frac{C R}{2}\equiv m_B$.

Let us emphasize that this metric is valid only outside some radius $r_B$ where the baryonic mass density vanishes. Thus, the singularity at $r=0$ is not practically relevant.

\subsection*{Circular velocity:}

The circular velocity  of a massive test particle moving in a static, spherically symmetric geometry  (at the equatorial plane $\theta = \frac{\pi}{2}$) is given by: $v^2 (r) = \frac{r f'}{2f}$ \cite{nuc}. For the spacetime solution (\ref{g-metric}) found above, we obtain:
\begin{eqnarray}
    v^2 (r)~ =~ \frac{\frac{\alpha}{(1-\alpha)^2} +\frac{1}{2}(1-3\alpha) \hspace{1mm} C \left(\frac{r}{R}  \right)^{-1+\alpha}}{\frac{1}{(1-\alpha)^2} - C \left(\frac{r}{R}  \right)^{-1+\alpha}}
\end{eqnarray}
Let us observe that for sufficiently large distances (far away from the luminous mass distribution), the second term above in the numerator and denominator becomes negligible. This leads to a constant circular velocity at large radii from the galactic centre: $v^2(r) \rightarrow \alpha$ as $r \rightarrow \infty$. Thus, the small parameter $\alpha$ acquires a physical interpretation in this large distance limit. 

In the weak-field limit ($C<<1$), let us expand the above expression for a suficiently large $r\lesssim R$:
\begin{eqnarray*}
v^2(r)~\approx~ \alpha+\Bigg[\frac{1}{2}(CR)(1-\alpha)^3 \Bigg]r^{-1}\Bigg(\frac{r}{R}\Bigg)^\alpha
\end{eqnarray*}
Since $\alpha<<1$, the leading ($\alpha$-independent) term above is $\frac{m_B}{r}$, which reproduces the Keplerian velocity.

\subsection*{Dynamical mass:}
The field equations (\ref{Req}) could be rewritten as below:
\begin{eqnarray}
R_{\alpha\beta}(\bar{\omega})-\frac{1}{2}g_{\mu\nu}R(\bar{\omega})=8\pi T^{(eff)}_{\mu\nu}
\end{eqnarray}
where the effective energy-momentum tensor $T^{(eff)}_{\mu\nu}$ originates from the scalar torsion and has no genuine matter content. 
The explicit expressions for the effective energy density and pressure components encoded by the above are given below:
\begin{eqnarray*}
   8\pi \rho &=&  \frac{\alpha}{r^2}\left[ -\frac{(2-\alpha)}{(1-\alpha)^2 } + C \left( \frac{R}{r} \right)^{1-\alpha} \right] \label{energydensity} \\
    8\pi P_r &=& \frac{\alpha}{r^2} \left[ \frac{4-\alpha}{(1-\alpha)^2} - 3C \left( \frac{R}{r} \right)^{1-\alpha} \right] \label{radialpress} \\
    8\pi P_\theta &=& \frac{\alpha}{r^2} \left[ \frac{\alpha}{(1-\alpha)^2} +(2-3\alpha) C \left( \frac{R}{r} \right)^{1-\alpha}  \right] =8\pi P_\phi\label{thetapress}
\end{eqnarray*}
Thus, the halo metric exhibits an anisotropic (effective) pressure.

 The total mass enclosed within a radius $r$ from the galactic centre is given by:
 \begin{eqnarray}\label{mass}
m(r)~&=&~4\pi \int dr ~r^2\rho(r)\nonumber\\
~&=&~-\frac{\alpha(2-\alpha)}{2(1-\alpha)^2}r~+~\frac{C R}{2}\Big(\frac{r}{R}\Big)^\alpha
\end{eqnarray}
We note that the leading term in the effective mass has a term linear in $r$, responsible for the (approximately) flat velocity profile at large distances. This geometric term has a negative sign. As is obvious though, this does not imply the existence of any matter with negative mass. The second term involves the baryon mass.


\section{Deflection of light}
As one of the most direct ways to understand the observational import of the galactic metric (\ref{g-metric}), we now investigate the deflection of a light ray propagating in it.
This spacetime is not asymptotically flat, which is a typical feature of any halo geometry. However, in reality, a galactic halo must terminate at some large but finite radius. Such a scenario could be realized by joining the galactic metric to an asymptotically flat metric at the boundary of the halo using the continuity of the metric components. The detailed analysis of the resulting deflection of a light ray propagating is presented in the Appendix. 

There is another possibility, that is, the halo could extend upto a very large (`infinite') radial distance. In that case a light ray coming from and going to infinity essentially perceives only the halo geometry, which completely determines the amount of gravitational bending. We address this problem below.

  The total angle of deflection of the ray is given by: $\delta = |2\Delta\phi - \pi|$, where $\Delta \phi$ is angular distance along a null geodesic in going from the radial distance at the closest approach ($r_0$) from the centre of the halo to the observer (at the equatorial plane $\theta=\frac{\pi}{2}$): 
\begin{eqnarray}
    \Delta \phi &=& \int_{r_0}^\infty dr \hspace{1mm} \frac{d\phi}{dr} = \int_{r_0}^\infty \frac{dr}{r} \left[ \frac{E^2}{L^2} \frac{r^2}{fg} - \frac{1}{g} \right]^{-\frac{1}{2}} \label{deflangle}
\end{eqnarray}
In the above, $E,L$ are the conserved quantities associated with the $t$ and $\phi$ motions, respectively, satisfying $\frac{E^2}{L^2} = \frac{f(r_0)}{r_0^2}$.


Next, using the identity:
\begin{eqnarray*}
   && \frac{E^2}{L^2} \frac{r^2}{fg} - \frac{1}{g} ~=~ \frac{1}{(1-\alpha)^2} \left[ \left( \frac{r}{r_0} \right)^{2-2\alpha} -1 \right]\times\nonumber\\
   && \left[ 1 - (1-\alpha)^2  \hspace{1mm} C \left( \frac{r}{R} \right)^{-1+\alpha} \frac{\left( \frac{r}{r_0} \right)^{2-2\alpha} + \left( \frac{r}{r_0} \right)^{1-\alpha} + 1}{\left( \frac{r}{r_0} \right)^{1-\alpha} +1} \right], \label{identity}
\end{eqnarray*}
we may evaluate the integral (\ref{deflangle}) in the weak-field limit.
Assuming $r_0\lesssim R$, the deflection of light passing through the infinitely extended halo is given by: 
\begin{eqnarray*}
   && \Delta \phi \approx \int_{r_0}^{\infty} \frac{dr}{r} (1-\alpha) \left[ \left( \frac{r}{r_0} \right)^{2-2\alpha} -1\right]^{-\frac{1}{2}}\Bigg[ 1 + \frac{(1-\alpha)^2}{2} C \left( \frac{r}{R} \right)^{-1+\alpha} \frac{\left( \frac{r}{r_0} \right)^{2-2\alpha} + \left( \frac{r}{r_0} \right)^{1-\alpha} + 1}{\left( \frac{r}{r_0} \right)^{1-\alpha} + 1} \Bigg] \nonumber \\
    &&\approx~ \frac{\pi}{2} ~+ ~\frac{2(1-2\alpha) m_B}{r_0} 
    \end{eqnarray*}
    where we have displayed only the leading order correction in the small parameter $\alpha$.
 The deflection angle thus is given by:
\begin{eqnarray}
    \delta~ \approx~ \frac{4 m_B}{r_0}~ -~ \frac{8 \alpha m_B}{r_0}
\end{eqnarray}
 The second term, reflecting the leading correction to the Einsteinian bending, is negative in sign and hence diminishes the bending. This is a new feature, and may be contrasted with the typical CDM or modified gravity models. This effect is suppressed to the Einsteinian contribution by a factor of $\alpha$, which is typically $\sim 10^{-6}$ for spiral galaxies. 
 
 The above result may be contrasted with the case of the galactic spacetime solution found earlier for the phase $\xi=0$ in ref.\cite{seng}, where the correction to the general relativistic deflection is found to be constant: $\delta - \frac{4 m_B}{r_0}=\frac{3}{2}\pi \alpha$ and is opposite in sign. In general, the result here has no analogue in the typical `dark matter' models, where the bending angle is enhanced \cite{blandford,harko1}, the exact modification being dependent on the density profile. For instance, the singular isothermal cold dark matter model with $\rho\sim \frac{1}{r^2}$ yields $\delta - \frac{4 m_B}{r_0}=2\pi \alpha$. Thus, the gravitational deflection of light allows a clear distinction of the galactic geometry found here from other viable alternatives from an observational perspective.

\section{Conclusions}

Based on a gravity action with a varying Newton's coupling, we have found a new set of exact solutions as a potential realization of galactic spacetimes. As the critical feature, these spacetimes correspond to rotation curves that approach a flat profile at large distances. These involve no matter source. Rather, the small spatial variation of the coupling contributes to a nontrivial scalar torsion that gets reflected through the equations of motion, essentially different from Einstein's field equations for spherical symmetry. In other words, the solutions obtained here have no analogue in Einstein gravity with or without standard matter.

It is worthwhile to contrast our results here with the galactic spacetime solutions for the other vacuum phase $\xi=0$, as reported in ref.\cite{seng}. While both correspond to (almost) flat profiles of the associated galactic rotation curves far away from the luminous mass and exhibit anisotropic effective pressure, there are three important differences. Firstly, the effective energy momentum tensor originating from the scalar torsion here has a non-negative trace unlike the latter where it is traceless. Secondly, the effective `mass' contains a negative nonbaryonic contribution. However, this does not imply the presence of a genuine negative mass, since there is no matter content. This result is in line with an earlier one where negative energy densities are shown to emerge necessarily in static, spherically symmetric configurations sourced by genuine scalar matter \cite{morales}.
Finally, the deflection of light in the limit of a very large halo radii receives a small correction that diminishes the bending angle. This provides a useful contrast with existing geometric frameworks or `dark-matter' models conceived to explain galactic rotation curves.

It is intriguing to observe that while the small spatial variation of the Newton's coupling is shown here to lead to  galactic metric solutions exhibiting asymptotically flat rotation curves, its temporal variation on the other hand suggests a potential solution to the coincidence problem as well as explain the acceleration of the universe \cite{sengcosm}. This provides a good reason to believe that further investigations into the dynamical consequences of the Lagrangian framework with varying gravitational couplings could be worthwhile.

\acknowledgments 
S. S. gratefully acknowledges the support (in part) of the ANRF (previously SERB), Govt. of India, through the MATRICS grant MTR/2021/000008.
 
 \bibliography{flat-rc-torsion-arxiv}

\appendix*

\section{Deflection of light propagating through a halo of finite radius}

Let us assume that the asymptotic observer's spacetime is flat. To be explicit, we choose this geometry to be Schwarzschild, which turns out to be an asymptotically flat solution of the equations of motion as already elucidated in Sec-III. 
Thus, the vacuum spacetime geometry is given by the halo metric (\ref{g-metric}) at $r\leq R_H$ and by the Schwarzschild exterior metric (with a corresponding mass $M$) at $r\geq R_H$, where $R_H$ is the halo boundary. The continuity of the two metrics at the boundary implies:
\begin{eqnarray}
R_H=R;~ C=\frac{2M}{R_H}+\frac{\alpha(2-\alpha)}{(1-\alpha)^2}~.
\end{eqnarray}
The continuity of the field $\xi(r)=\frac{1}{C_1^2}\Big(\frac{R}{r}\Big)^{2\alpha}$ implies that $\xi(R_H)=\frac{1}{C_1^2}$, which is positive, defines the Newton's constant (set to $1$ here) characterizing the Schwarzschild geometry.

Note that here we have: $\frac{E^2}{L^2} = \frac{f(r_0)}{r_0^2} = \frac{f(r_*)}{r_*^2}$ ($r_*$ denotes the distance at the closest approach in the Schwarzschild geometry). We shall evaluate the above integral under the weak-field approximation, namely, $C<<1, ~\frac{M}{r_*}<<1$. 

The contribution to the integral due to the passage of the light ray within the halo reads:
\begin{eqnarray}
    && \Delta \phi_1 \approx \int_{r_0}^{R_H} \frac{dr}{r} (1-\alpha) \left[ \left( \frac{r}{r_0} \right)^{2-2\alpha}-1 \right]^{-\frac{1}{2}} \times \nonumber\\
    &&\left[ 1 - \frac{(1-\alpha)^2}{2} C \left( \frac{r}{R} \right)^{-1+\alpha} \frac{\left( \frac{r}{r_0} \right)^{2-2\alpha} + \left( \frac{r}{r_0} \right)^{1-\alpha} + 1}{\left( \frac{r}{r_0} \right)^{1-\alpha} + 1} \right] \nonumber \\
    &&=~ \tan^{-1} \left[ \left(\frac{R}{r_0}\right)^{2-2\alpha} - 1 \right]^{\frac{1}{2}} ~+~ \frac{(1-\alpha)^2 C}{2}  \left[ \left( \frac{R}{r_0} \right)^{2-2\alpha} - 1 \right]^{\frac{1}{2}} \Bigg[\frac{2 \left( \frac{R}{r_0} \right)^{1-\alpha} + 1}{\left( \frac{R}{r_0} \right)^{1-\alpha} + 1}\Bigg]
\end{eqnarray}

The remaining contribution due to the propagation in the Schwarzschild geometry is given by:
\begin{eqnarray}
    &&\Delta \phi_2 \approx \int_{R_H}^\infty \frac{dr}{r} \left( \frac{r^2}{r_*^2} -1 \right)^{-\frac{1}{2}} \left[ 1 + \frac{M}{r} + \frac{M r}{r_* (r+ r_*)} \right] \nonumber\\
    &&=~ \frac{\pi}{2} ~+~ \frac{2M}{r_*}~-~ \tan^{-1} \left[ \left( \frac{R}{r_*} \right)^2 - 1 \right]^{\frac{1}{2}}  ~-~ \frac{M}{R} \left[ \left( \frac{R}{r_*} \right)^2 -1 \right]^{\frac{1}{2}} \left( \frac{\frac{2R}{r_*} + 1}{\frac{R}{r_*} + 1} \right)
\end{eqnarray}
As a result, the total deflection angle upon using the expression for $C$ reads:
\begin{eqnarray}
    && \delta ~=~  \frac{4M}{r_*} + 2\tan^{-1} \left[ \left(\frac{R}{r_0}\right)^{2-2\alpha} - 1 \right]^{\frac{1}{2}}\nonumber\\
    && +~ (1-\alpha)^2 \left[ \frac{2M}{R} + \frac{\alpha(2-\alpha)}{(1-\alpha)^2} \right]  \left[ \left( \frac{R}{r_0} \right)^{2-2\alpha} - 1 \right]^{\frac{1}{2}} \frac{2 \left( \frac{R}{r_0} \right)^{1-\alpha} + 1}{\left( \frac{R}{r_0} \right)^{2-2\alpha} + 1} \nonumber \\
    &&-~ 2 \tan^{-1} \left[ \left( \frac{R}{r_*} \right)^2 - 1 \right]^{\frac{1}{2}} ~-~ \frac{2M}{R} \left[ \left( \frac{R}{r_*} \right)^2 -1 \right]^{\frac{1}{2}} \left( \frac{\frac{2R}{r_*} + 1}{\frac{R}{r_*} + 1} \right)
\end{eqnarray}
For $r_0 = r_* = R$, corresponding to a light ray passing only through the Schwarzschild geometry without entering the halo, we recover the Einsteinian deflection: $\delta = \frac{4M}{r_*}$. The general result obtained above is characterized completely by the ratio $\frac{R}{r_0} \approx \frac{R}{r_*}$, given the asymptotic rotation velocity. The deflection of a light ray entering only the outer region of the halo where the rotation velocity becomes flat corresponds to the case $C = 0$, leading to:
\begin{eqnarray}
   && \delta ~\approx~ \frac{4M}{r_0} + 2\tan^{-1} \left[ \left(\frac{R}{r_0}\right)^{2-2\alpha} - 1 \right]^{\frac{1}{2}}~-~ 2 \tan^{-1} \left[ \left( \frac{R}{r_0} \right)^2 - 1 \right]^{\frac{1}{2}} \nonumber\\
   &&-~ \frac{2M}{R} \left[ \left( \frac{R}{r_0} \right)^2 -1 \right]^{\frac{1}{2}} \left( \frac{\frac{2R}{r_0} + 1}{\frac{R}{r_0} + 1} \right)
\end{eqnarray}

\end{document}